\documentclass{osa-article}
\journal{oe}

\articletype{Research Article}

\begin{document}

\title{Nonperturbative solution to the integral equation of scattering theory}

\author{Brian Slovick\authormark{*} and Srini Krishnamurthy}

\address{Applied Sciences Division SRI International, Menlo Park California 94025, USA}

\email{\authormark{*}brian.slovick@sri.com} 

\begin{abstract}
We obtain a nonperturbative, analytical solution to integral equation of scattering theory by assuming the field within the scattering object is a spherical wave with a scattering amplitude equal to that of the far field. This approximation transforms the integral equation into a simple algebraic equation which can be readily solved to obtain a closed-form expression for the scattering amplitude. We show this approximation is valid for homogeneous potentials of compact support, namely circular and square cylinders, and that the calculated scattering cross sections for spheres and square cylinders are accurate for frequencies through the fundamental resonance. Then we apply our analytical expression to the inverse scattering problem for spheres and show that accurate reconstructions are possible even under resonance conditions. The simplicity and accuracy of our method suggest it can be a reliable and efficient tool for understanding a wide range of scattering problems in optics.
\end{abstract}

\section{Introduction}
Direct and inverse scattering problems are prevalent throughout physics \cite{Newton2013,Behura2014,Slovick2014,Bao2015,Levinson2016I,Ablowitz2016}. In the direct scattering problem, the properties of the object are used to determine the scattered fields, a scenario routinely encountered in particle physics \cite{Taylor2006,Chadan2012}, acoustics \cite{Colton2000}, and electromagnetics \cite{Jacob2006,Colton2012,Slovick2014,Papasimakis2016}. In the inverse scattering problem \cite{Colton2012,Levinson2016I,Levinson2016II,Isakov2017}, the properties of the object are determined from the scattered fields. Inverse scattering is important for applications such as remote sensing \cite{Colton2003,Leone2003,Woodhouse2017}, medical imaging \cite{Ralston2006,Marks2007}, and design and optimization of complex materials \cite{Alu2005,Fiddy2006,Fiddy2007,Hakansson2007,Alexopoulos2007,Alu2008,Donato2017,Cheng2017}. A general analytical solution to the direct and inverse scattering problem would benefit these fields and applications by providing provide new physical insight and more efficient computation.

The predominant methods for solving the scattering problem are partial wave analysis \cite{Landau1996,Newton2013}, the Born approximation \cite{Devaney1982,Devaney1983,Colton2000,Taylor2006}, and numerical methods \cite{Vainikko2000,Newton2013,Isakov2017}. All have their advantages and limitations. Partial wave analysis is exact, but is limited to objects with rotational symmetry and cannot be used to solve the inverse scattering problem because the equations are transcendental \cite{Newton2013,Colton2012}. The Born approximation provides a closed-form expression and is valid for all object shapes \cite{Wolf1999,Colton2000}, but is limited to weak scattering conditions \cite{Devaney1982,Devaney1983,Fiddy2015}. Numerical methods are exact within numerical error, but provide limited physical insight and are inefficient due to the difficulty of solving integral equations \cite{Vainikko2000,Newton2013,Isakov2017}.

In this Letter, we obtain a nonperturbative closed-form solution to the integral equation of scattering theory using what we call the extended far-field approximation (EFA). The EFA assumes the field within the object has the same scattering amplitude as the far-field scattering amplitude. Using full-wave simulations, we show the EFA is valid well into the scattering volume of homogeneous potentials of compact support, namely circular and square cylinders, and that the scattering cross section for spheres and square cylinders obtained with EFA agrees well with exact calculations obtained by partial wave analysis, for frequencies through the fundamental resonance. In addition, assuming convergence of an infinite geometric series, our closed-form expression can be inverted to obtain the potential in terms of the scattered field. Within the limits of this approximation, the EFA provides much improved inverse scattering reconstructions for compact spherical objects compared to the Born approximation. The simplicity and accuracy of our method suggest it can be a reliable and efficient tool for understanding a wide range of scattering problems in optics.

\section{Approach}
Our closed-form expression can be derived from the inhomogeneous wave equation for a time-harmonic scalar field $U(\textbf{r},\omega)$ \cite{Landau1996,Wolf1999},
\begin{equation}
(\nabla^2+k^2)U(\textbf{r},\omega)=-4 \pi F(\textbf{r},\omega)U(\textbf{r},\omega),
\end{equation}
where $F(\textbf{r},\omega)$ is the scattering potential and $k$ is the wavenumber. With the Sommerfeld radiation condition, the solution to Eq. (1) is given by the sum of the incident field $U_i(\textbf{r},\omega)$ and the scattered field \cite{Landau1996,Wolf1999}
\begin{equation}
U_s(\textbf{r},\omega)=\int d^3r' G(\mathbf{r}-\mathbf{r}',\omega) F(\textbf{r}',\omega) [U_i(\mathbf{r}',\omega)+U_s(\mathbf{r}',\omega)],
\end{equation}
where the Green's function
\begin{equation}
G(\textbf{r}-\textbf{r}',\omega)=\frac{e^{ik|\textbf{r}-\textbf{r}'|}}{|\textbf{r}-\textbf{r}'|},
\end{equation}
is the solution to $(\nabla^2+k^2)G(\textbf{r}-\textbf{r}',\omega)=-4 \pi \delta^{(3)}(\textbf{r}-\textbf{r}')$. In the far-field approximation, $|\textbf{r}-\textbf{r}'|\approx r-\frac{\textbf{k}_s}{k} \cdot \textbf{r}'$, where $\textbf{k}_s$ is a vector of magnitude $k$ in the direction of $\textbf{r}$, and the Green's function simplifies to
\begin{equation}
G(\textbf{r}-\textbf{r}',\omega)\approx \frac{e^{ikr}}{r}e^{-i \textbf{k}_s \cdot \textbf{r}'  }.
\end{equation}
Noting that $U_s(\textbf{r},\omega)=\frac{e^{ikr}}{r}f(\textbf{k}_s,\textbf{k}_i)$, where $f(\textbf{k}_s,\textbf{k}_i)$ is the scattering amplitude, for an incident plane wave of the form $U_i(\textbf{r}',\omega)=e^{\textbf{k}_i \cdot \textbf{r}'}$, Eq. (2) reduces to
\begin{equation}
f(\textbf{k}_s,\textbf{k}_i)= \int d^3r'  F(\textbf{r}',\omega) e^{-i(\textbf{k}_s-\textbf{k}_i)\cdot \textbf{r}'}
+\int d^3r' F(\textbf{r}',\omega)e^{-i\textbf{k}_s\cdot \textbf{r}'}U_s(\textbf{r}',\omega).
\end{equation}
Equation (5) is the integral equation of scattering theory. It is generally difficult to solve because the scattered field inside the scattering volume is not known a priori. In the first Born approximation, $U_s(\textbf{r}',\omega)$ is set equal to zero, leading to a Fourier transform relationship between the scattering amplitude and the potential \cite{Devaney1982,Wolf1999}. For improved accuracy, the solution obtained with the first Born approximation (and subsequent solutions) can be used as the internal field, leading to the Liouville-Neumann series. However, this series only converges for small potentials, so it is not applicable to resonant structures, the focus of this work.

The EFA assumes the internal field is a spherical wave with the same scattering amplitude as the far-field scattering amplitude, i.e.,
\begin{equation}
U_s(\textbf{r}',\omega)=\frac{e^{ikr'}}{r'} f(\textbf{k}_s,\textbf{k}_i).
\end{equation}
This allows the second term on the right hand side of Eq. (5) to be factorized, leading to the following closed-form expression for the scattering amplitude:
\begin{equation}
f(\textbf{k}_s,\textbf{k}_i)=\frac{\int d^3r'  F(\textbf{r}',\omega) e^{-i(\textbf{k}_s-\textbf{k}_i)\cdot \textbf{r}'} }{1-\int d^3r'  F(\textbf{r}',\omega) e^{-i\textbf{k}_s \cdot \textbf{r}'} \frac{e^{ikr'}}{r'} }.
\end{equation}
Equation (7) gives the scattering amplitude as a function of the potential, and in principle, is valid for arbitrary potentials. For small scattering potentials, the denominator is one and Eq. (7) reduces to the well-known expression obtained with the first-Born approximation.

In forward scattering problems, $F(\textbf{r}',\omega)$ is known and Eq. (7) can be used to calculate the scattering amplitude. Alternatively, in inverse scattering problems, $f(\textbf{k}_s,\textbf{k}_i)$ is known and $F(\textbf{r}',\omega)$ can be determined. Defining the momentum transfer as $\textbf{q}\equiv \textbf{k}_s-\textbf{k}_i$ and taking the Fourier transform of Eq. (7) with respect to $\textbf{q}$, we obtain
\begin{eqnarray}
F(\textbf{r},\omega)=F_B(\textbf{r},\omega) \left[ 1-\int d^3r'  F(\textbf{r}',\omega) e^{-i\textbf{k}_s \cdot \textbf{r}'} \tfrac{e^{ikr'}}{r'} \right], \quad \text{where}
\end{eqnarray}
\begin{eqnarray}
F_B(\textbf{r},\omega)\equiv \frac{1}{(2\pi)^3} \int d^3 q e^{i\textbf{q} \cdot \textbf{r} } f(\textbf{k}_s,\textbf{k}_i)
\end{eqnarray}
is the potential obtained from the first Born approximation. For elastic scattering $|\textbf{q}|\le 2k$, and the potential from Eq. (9) is a low-pass filtered version of the potential \cite{Vainikko2000,Colton2012,Fiddy2015}. The lack of high spatial frequencies is why inverse scattering based on far-field data is not unique \cite{Bleistein1977,Devaney1978}, as objects with different subwavelength structure give rise to the same scattered far fields. By iteratively substituting $F(\textbf{r},\omega)$ in Eq. (8) and applying the infinite geometric series sum, we obtain
\begin{eqnarray}
F(\textbf{r},\omega)= \frac{F_B(\textbf{r},\omega)}{1+\int d^3r'  F_B(\textbf{r}',\omega) e^{-i\textbf{k}_s \cdot \textbf{r}'} \frac{e^{ikr'}}{r'} }.
\end{eqnarray}
The closed-form expressions in Eq. (7) and (10) form the basis for direct and inverse scattering studies, respectively. Their accuracy depends on the validity of the EFA and the requirement that the absolute value of the integral in the denominator of Eq. (10) is less than 1 for series convergence.

\section{Validation}
To demonstrate the validity of our approach, we consider scalar scattering with $F(\textbf{r},\omega)=k^2[\epsilon(\textbf{r},\omega)-1]/4\pi$, where $\epsilon(\textbf{r},\omega)$ can be considered the relative permittivity for scalar fields. Using full-wave electromagnetic simulations, we calculated the scattering amplitude for infinite cylinders of radius $a$ and frequency-independent homogeneous permittivity $\epsilon_r$. To compare these results to our scalar-wave formalism, we consider light polarized along the axis of the cylinder. The calculated scattering amplitude at different distances from the center of cylinders with $\epsilon_r=2$ (top), $\epsilon_r=6$ (middle), and $\epsilon_r=10$ (bottom) and diameters of $2 a k\sqrt{\epsilon_r}$=1 (left), 2 (middle), and 3 (right) are shown in Fig. 1. Consistent with the EFA, we find that the scattering amplitude inside the cylinders is very close to the far-field values essentially for all cases except very close to the origin ($r<0.5a$). To confirm the validity of EFA is not limited to scatterers with circular symmetry, we carried out a similar study for infinite square rods of side length $2w$. The calculated scattering amplitude at different distances from the rods are shown in Fig. 2. We find that the EFA is equally valid for this case for distances of $r>0.5w$.

\begin{figure}
\centering
\includegraphics[width=120mm]{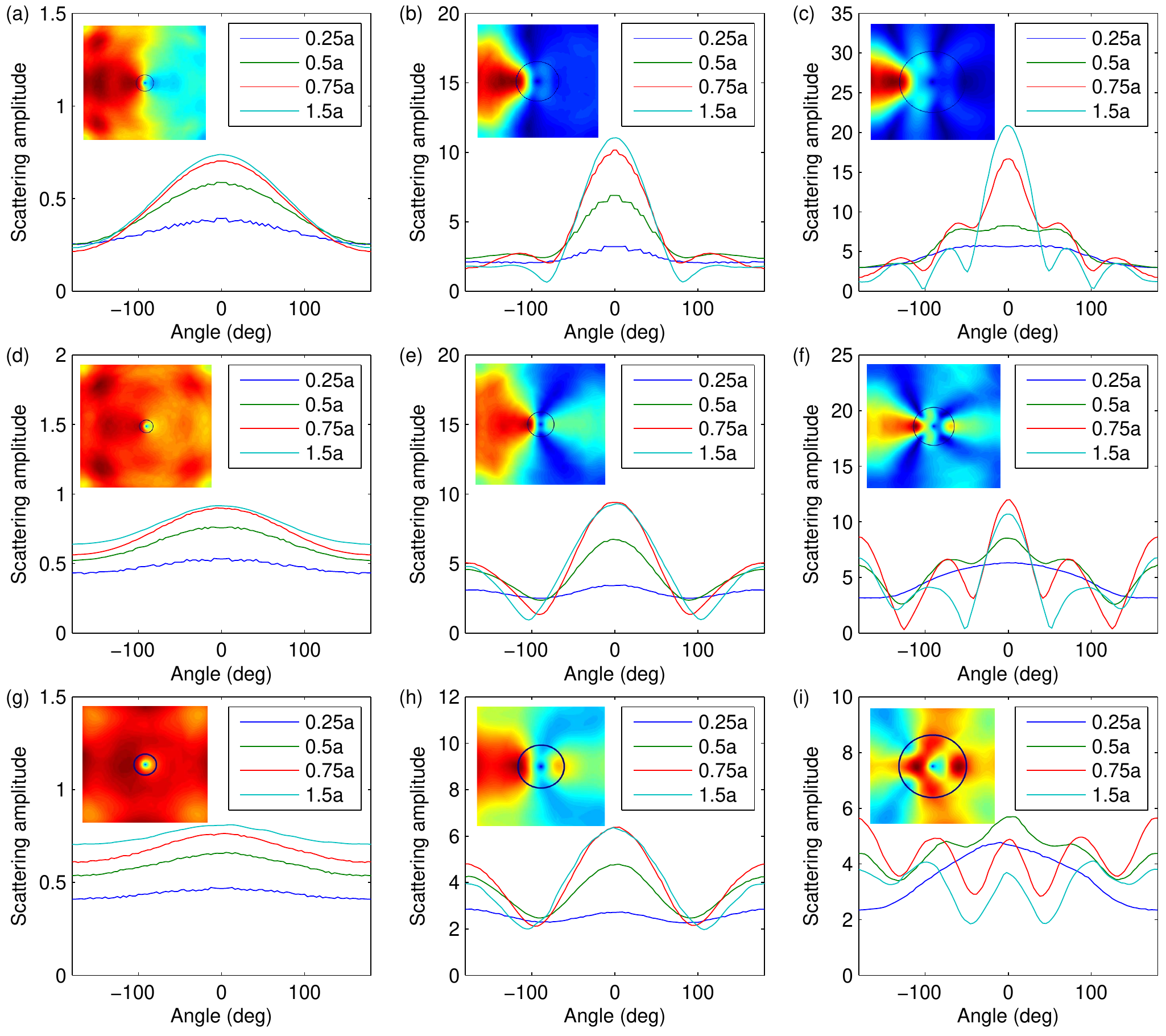}
\caption{Calculated scattering amplitude at different distances from the center of dielectric cylinders with $\epsilon_r=2$ (top), $\epsilon_r=6$ (middle), and $\epsilon_r=10$ (bottom) and diameters of $2 a k\sqrt{\epsilon_r}$=1 (left), 2 (middle), and 3 (right).}
\end{figure}

\begin{figure}
\centering
\includegraphics[width=120mm]{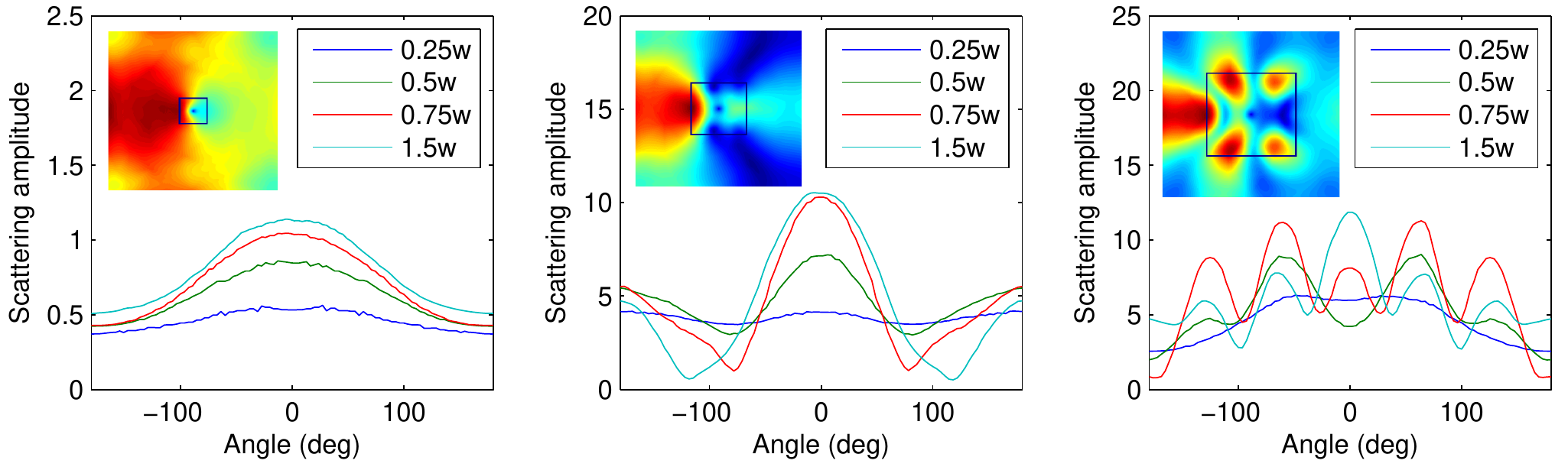}
\caption{Calculated scattering amplitude at different distances from the center of infinitely long square rods with $\epsilon_r=6$ and side lengths of $2w k\sqrt{\epsilon_r}$=1 (left), 2 (middle), and 3 (right).}
\end{figure}

\begin{figure}
\centering
\includegraphics[width=120mm]{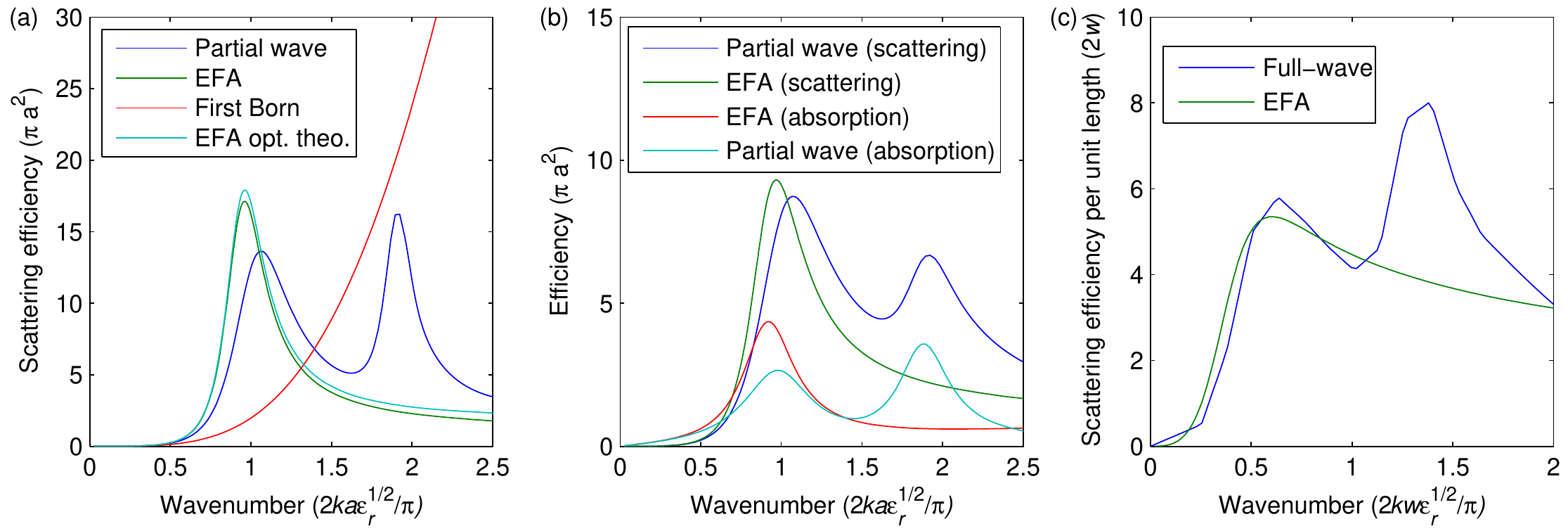}
\caption{ Scattering and absorption cross sections, in units of the cross sectional area, as a function of wavenumber calculated using partial-wave analysis, EFA, and the first-Born approximation for spheres with (a) $\epsilon_r=10$, and (b) $\epsilon_r=10+1i$. The cross section obtained by applying the optical theorem to the EFA scattering amplitude is also shown. (c) Scattering efficiency per unit length for infinite square cylinders with $\epsilon_r=10$ and width $2w$.}
\end{figure}

\section{Direct Scattering}
Having confirmed the validity and limitations of the EFA, we used Eq. (7) to calculate the scattering amplitude and cross section of scalar waves for spheres with $\epsilon_r=10$ and compared the results to the exact solution obtained by partial wave analysis, i.e., decomposing the wave function into spherical harmonics and imposing continuity of the field and its derivative. For spheres, the EFA leads to the following closed-form expression for the scattering amplitude:
\begin{equation}
f(q,k)=\frac{k^2}{q^3}(\epsilon_r-1)\frac{\sin{qa}-qa\cos{qa}}{1+\tfrac{1}{4}(\epsilon_r-1)(e^{i2ka}-i2ka-1)}
\end{equation}
Figure 3a shows the scattering cross section as a function of wavenumber (divided by the diameter $2a$) for lossless spheres with $\epsilon_r=10$, obtained by integrating Eq. (13) over angle as $2\pi\int_0^{2k} dq \tfrac{q}{k^2} |f(q,k)|^2$ (green). Also shown is the exact cross section calculated using partial wave analysis (blue). The values calculated using the first Born approximation are shown in red. For small wavenumbers, or long wavelengths, corresponding to Rayleigh scattering, all three calculations are in agreement. As the wavenumber increases, the Born method remains valid for $2ka\sqrt{\epsilon_r}/\pi<0.4$, while the EFA remains valid for $2ka\sqrt{\epsilon_r}/\pi<1.6$. In addition, the EFA accurately predicts both the location of the fundamental resonance around $2ka\sqrt{\epsilon_r}/ \pi=1$, corresponding to the isotropic monopole resonance, and its magnitude. On the other hand, the Born approximation fails near resonance, predicting an increasing cross section for all wavenumbers. The second resonance is not predicted with EFA, consistent with the results shown in Fig. 1. Also shown is the cross section obtained with EFA calculated using the optical theorem as $\tfrac{4\pi}{k} \text{Im} f(0,k)$ (cyan). We find excellent agreement  with the cross section obtained by angle integration, confirming the EFA satisfies the optical theorem for this case. Figure 3b shows the results for lossy spheres with $\epsilon_r=10+1i$. In this case the absorption cross sections are calculated as the difference between the scattering cross section and the total cross section obtained from the optical theorem. Despite the relatively large amount of absorption, the EFA compares well with the partial wave results.

To investigate the accuracy of the EFA for objects with non-circular symmetry, we calculated the scattering efficiency (per unit length) for infinite square cylinders with $\epsilon_r=10$ and width $2w$ [Fig. 3(c)]. Similar to spheres, we find the EFA accurately captures the fundamental resonance, both the amplitude and location, while the higher-order resonances are again absent. The EFA remains accurate for $2kw \sqrt{\epsilon_r}/\pi<1$, consistent with the results of Fig. 2.

\begin{figure}
\centering
\includegraphics[width=80mm]{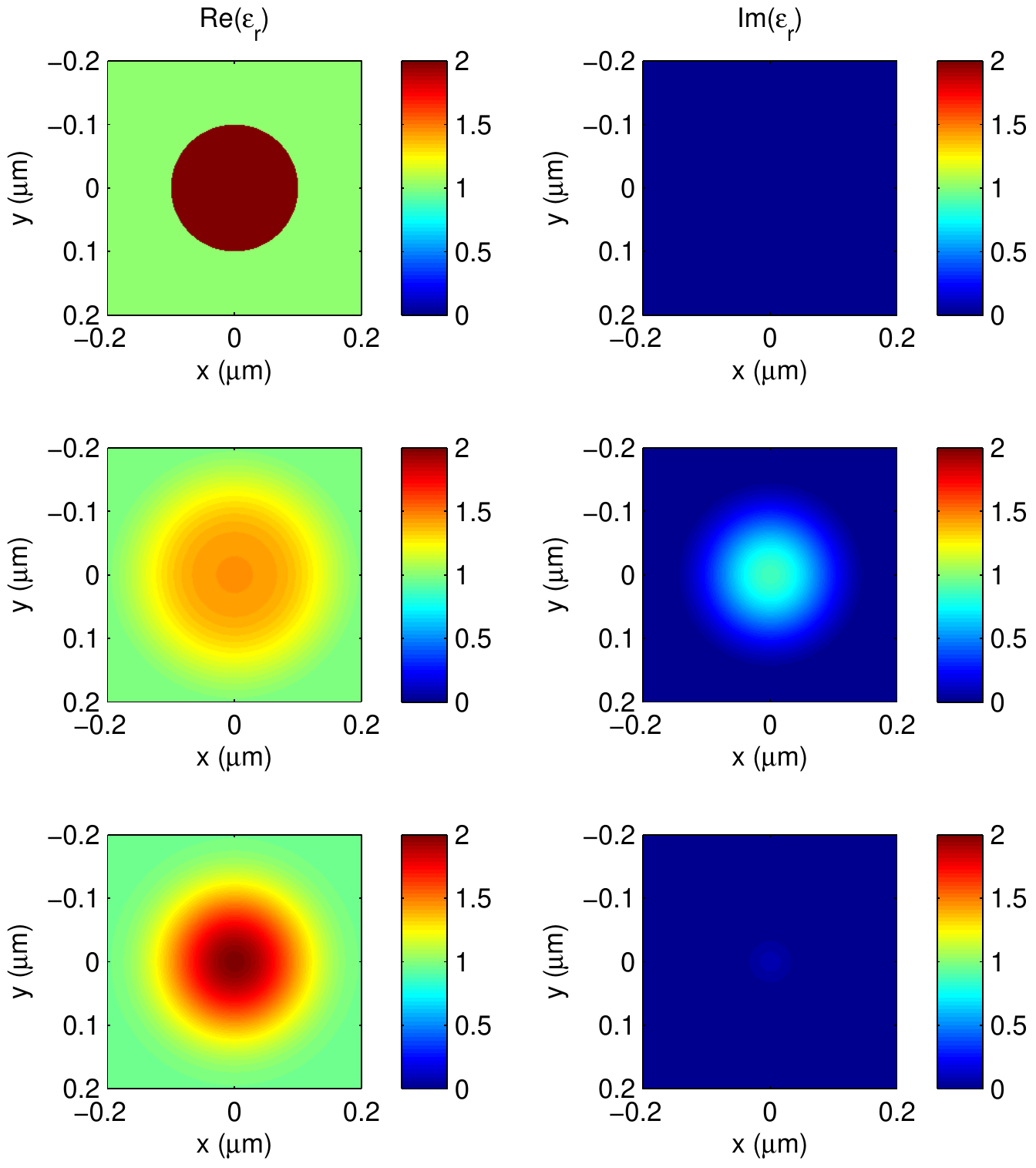}
\caption{Real (left panel) and imaginary (right panel) part of $\epsilon_r$ used in direct scattering (top panel) and reconstructions obtained by inverse scattering using first Born (middle panel) and EFA (bottom panel) methods for $\epsilon_r=2$.}
\end{figure}

\begin{figure}
\centering
\includegraphics[width=80mm]{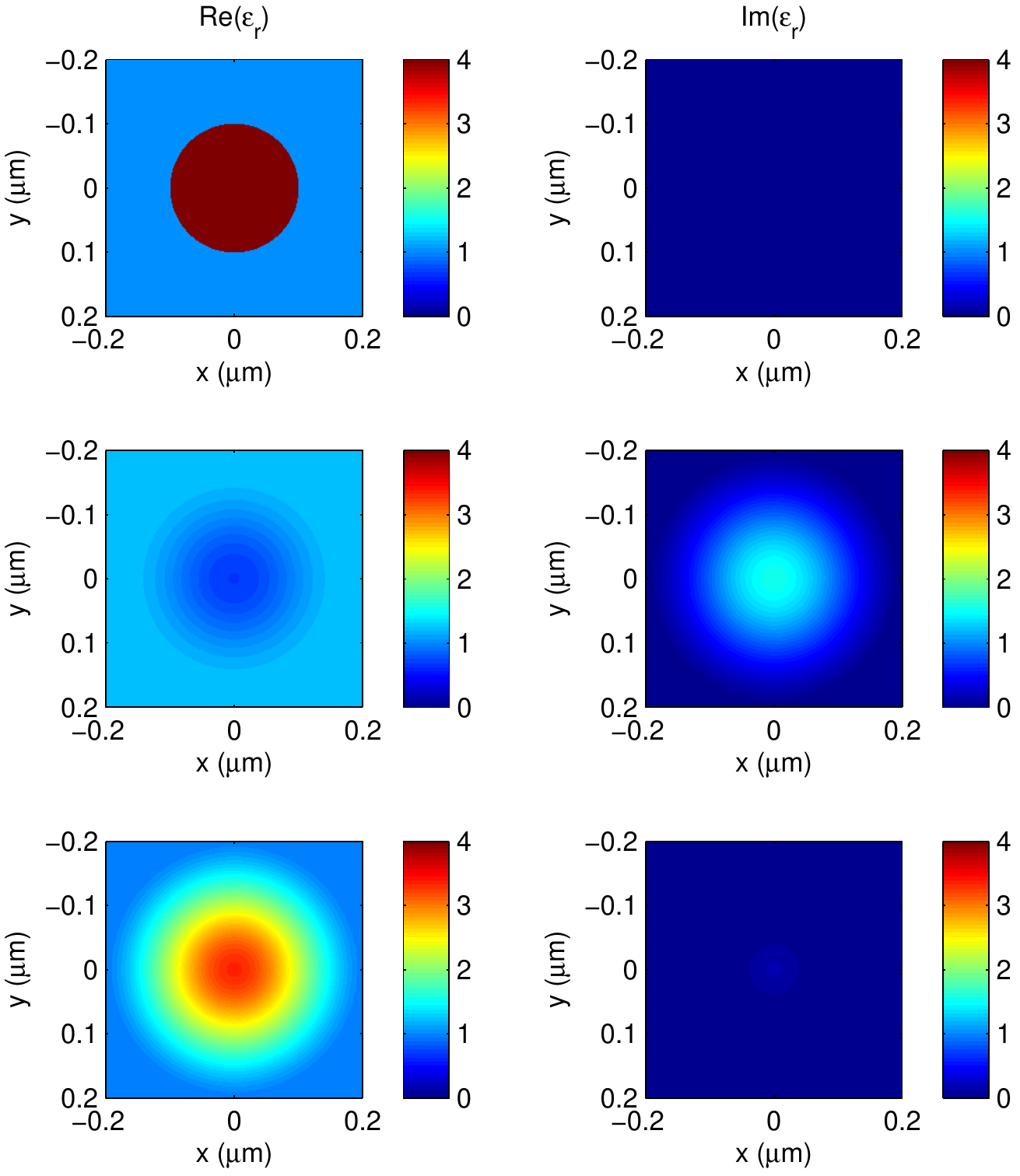}
\caption{Real (left panel) and imaginary (right panel) part of $\epsilon_r$ used in direct scattering (top panel) and reconstructions obtained by inverse scattering using first Born (middle panel) and EFA (bottom panel) methods for $\epsilon_r=4$.}
\end{figure}

\begin{figure}
\centering
\includegraphics[width=80mm]{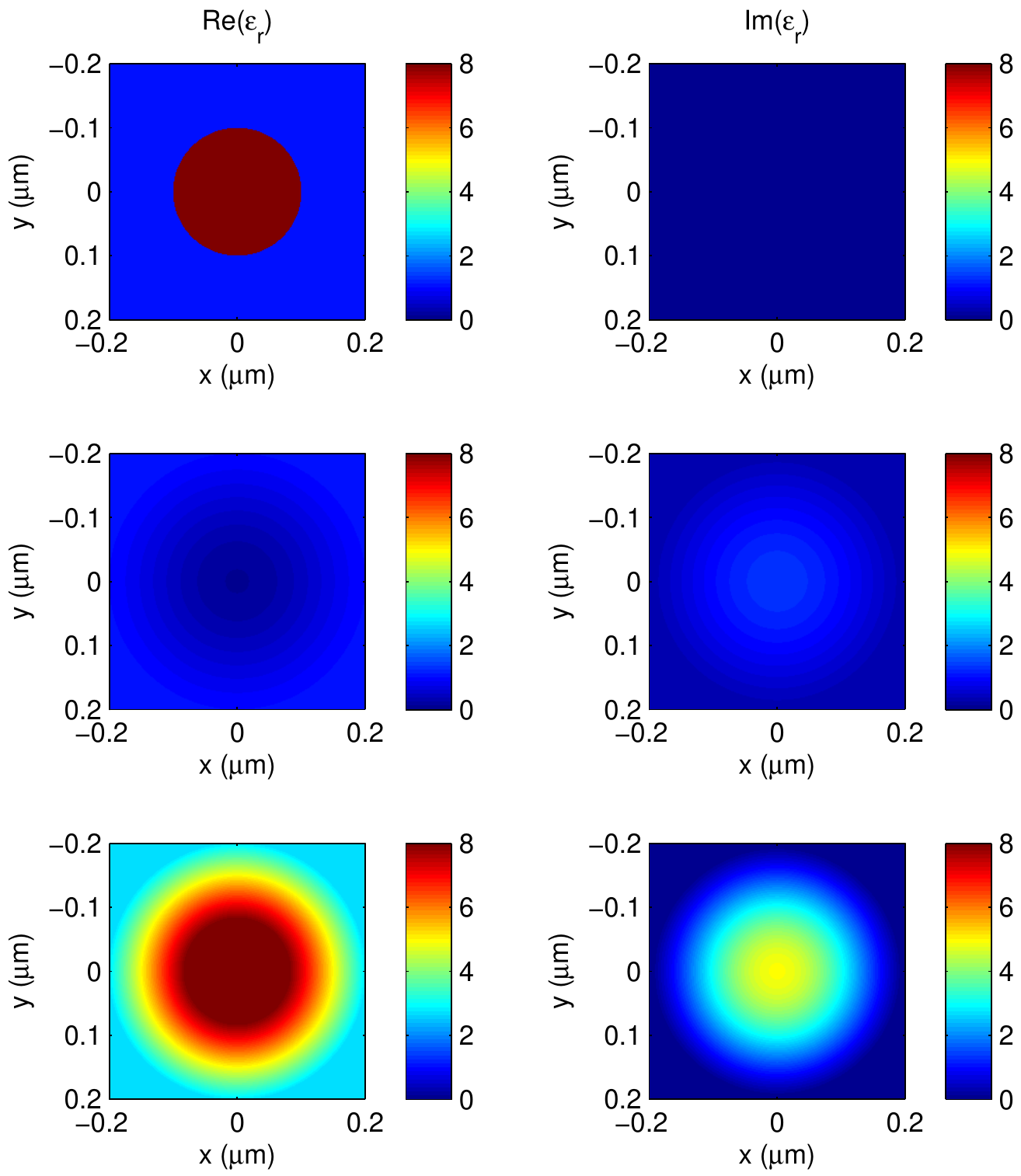}
\caption{ Real (left panel) and imaginary (right panel) part of $\epsilon_r$ used in direct scattering (top panel) and reconstructions obtained by inverse scattering using first Born (middle panel) and EFA (bottom panel) methods for $\epsilon_r=8$.}
\end{figure}

\section{Inverse Scattering}
The analytical expression in Eq. (10) can be used to perform inverse scattering reconstructions. For validation we consider spheres, allowing partial wave analysis to be used to obtain the scattering amplitude for all values of $q$. This result is then used in Eqs. (9) and (10) to obtain the permittivity reconstruction. As discussed earlier, the accuracy of the reconstruction is limited by the maximum spatial frequency in the scattered field, i.e., the limit of the integral in Eq. (9). To obtain the highest resolution, the shortest possible wavelength should be used. However, since the derivation of Eq. (10) requires convergence of a geometric series, we choose the wavelength for which the absolute value of the integral in the denominator is approximately equal to 1. Figures 5 through 7 compare the EFA-calculated real (left) and imaginary (right) parts of the permittivity distribution with the actual values and those obtained with the first Born approximation.

For $\epsilon_r=2$ (Fig. 4), the scattering is weak and the real part predicted using the first Born approximation agrees reasonably well with the actual values. However, the imaginary part is non-zero near the center. The EFA is considerably more accurate with respect to both the real and imaginary parts of $\epsilon_r$. The EFA reconstruction is slightly unresolved due to the limited spatial frequency used in the inversion. For $\epsilon_r=4$ (Fig. 5), the scattering is stronger and, expectedly, the prediction obtained with the Born approximation is poor, both the real and imaginary parts. The EFA continues to correctly predict purely real permittivity. The calculated values agree reasonably well with the actual values, though with less resolution compared to the previous case. Increasing the resolution would require a larger $q$ limit, but this would cause the series to diverge, leading to worse agreement. Despite this limitation, the EFA is reasonably accurate for this case. For $\epsilon=8$ (Fig. 6), the scattering is even stronger and again the Born approximation fails to predict the correct permittivity. Although the EFA predicts the real part and shape reasonably well, the imaginary part is unacceptably large. This is because the scattering potential, proportional to $\epsilon_r-1$, is large for this case, which limits the maximum value of $q$ that can be used before the denominator term in Eq. (10) approaches 1, in which case the series sum diverges. Thus, for this strong scattering regime, inverse scattering with EFA only provides qualitative predictions.

The accuracy of the EFA for homogeneous spheres can be understood from the field plots of Fig. 1. In regions where the EFA is invalid, near the center of the scatterer, the fields are close to zero, so their contribution to the scattered field, from Eq. (2), is small. This raises the question whether the EFA is valid for inhomogeneous potentials with large values near the origin. To investigate this case, we applied the EFA to calculate the scattering amplitude and cross section for the shielded Coulomb potential, also known as the Yukawa potential. We found the results did not satisfy the optical theorem, indicating a breakdown of the EFA. One possible way to resolve this is to relax the EFA such that the internal field is only proportional to the far-field scattering amplitude, rather than equal to it, and then adjust the proportionality constant to satisfy the optical theorem. Although this may increase the generality of the EFA, it increases the complexity significantly, and for this reason is not studied in detail here.

\section{Summary}
In summary, we obtained a nonperturbative closed-form solution to the integral equation of scattering theory by assuming the scattered field within the object is a spherical wave with a scattering amplitude equal to that of the far field. This approximation transforms the integral equation into an algebraic equation which can be readily solved to obtain the scattering amplitude. We found this approximation to be largely valid for homogeneous potentials of compact support, namely circular and square cylinders, and that the calculated scattering cross section for spheres and square cylinders agree well with exact results for frequencies through the fundamental resonance. The closed-form expression also enables reconstruction of the scattering object in inverse scattering studies. We applied this approach to reconstruct the permittivity profile of spheres and showed that it yields considerable improvement over the Born approximation. The simplicity and accuracy of our method suggest it can be a reliable and efficient tool for understanding a wide range of scattering problems in optics.

\section{Acknowledgments}
The authors are grateful to Dr. Michael Fiddy for several critical fruitful discussions on this topic and to DARPA for funding through Contract \# HR001117C0118.

\bibliography{bib}

\end{document}